\documentclass[a4paper, amsfonts, amssymb, amsmath, reprint, showkeys, nofootinbib, twoside,superscriptaddress]{revtex4-1}
\usepackage[english]{babel}
\usepackage[utf8]{inputenc}
\usepackage{amsthm}
\usepackage{mathtools}
\usepackage{physics}
\usepackage{xcolor}
\usepackage{graphicx}
\usepackage[left=23mm,right=13mm,top=35mm,columnsep=15pt]{geometry} 
\usepackage{adjustbox}
\usepackage{placeins}
\usepackage[T1]{fontenc}
\usepackage{lipsum}
\usepackage{csquotes}
\usepackage[colorinlistoftodos, color=green!40, prependcaption]{todonotes}
\usepackage{tcolorbox}
\usepackage{graphicx}
\usepackage{caption}
\usepackage{subcaption}
\usepackage[pdftex, pdftitle={Article}, pdfauthor={Author}]{hyperref} 
\usepackage{xcolor}

\begin{document}
\title{On Hawkes Processes with Infinite Mean Intensity}

\author{Cecilia Aubrun}
\affiliation{Chair of Econophysics and Complex Systems, \'Ecole polytechnique, 91128 Palaiseau Cedex, France}
\affiliation{LadHyX UMR CNRS 7646, \'Ecole polytechnique, 91128 Palaiseau Cedex, France}

\author{Michael Benzaquen}
\email{michael.benzaquen@polytechnique.edu}
\affiliation{Chair of Econophysics and Complex Systems, \'Ecole polytechnique, 91128 Palaiseau Cedex, France}
\affiliation{LadHyX UMR CNRS 7646, \'Ecole polytechnique, 91128 Palaiseau Cedex, France}
\affiliation{Capital Fund Management, 23 Rue de l’Universit\'e, 75007 Paris, France}
\author{Jean-Philippe Bouchaud}
\affiliation{Chair of Econophysics and Complex Systems, \'Ecole polytechnique, 91128 Palaiseau Cedex, France}
\affiliation{Capital Fund Management, 23 Rue de l’Universit\'e, 75007 Paris, France}
\affiliation{Académie des Sciences, 23 Quai de Conti, 75006 Paris, France}

\date{\today} 

\begin{abstract}
The stability condition for Hawkes processes and their non-linear extensions usually relies on the condition that the mean intensity is a finite constant. It follows that the total endogeneity ratio needs to be strictly smaller than unity.
In the present note we argue that it is possible to have a total endogeneity ratio greater than unity without rendering the process unstable. In particular, we show that, provided the endogeneity ratio of the linear Hawkes component is smaller than unity, Quadratic Hawkes processes are always stationary, although with infinite mean intensity when the total endogenity ratio exceeds one. This results from a subtle compensation between the inhibiting realisations (mean-reversion) and their exciting counterparts (trends).
\end{abstract}


\maketitle

\section{Introduction} \label{sec:intro}

Hawkes processes have been used in various fields to model endogenous dynamics, where past activity triggers more activity in the future. Indeed, Hawkes processes were found to be relevant to capture the self-excited nature of the dynamics in biological neural networks \cite{osorio2010epileptic,sornette2010prediction}, in financial markets \cite{bacry2015hawkes,fosset2021non}, in seismologic activity (earthquakes) \cite{hawkes1971spectra}, and also in crime rates or riot propagation \cite{mohler2011self,bonnasse2018epidemiological}. Standard linear Hawkes processes are basically akin to a branching process, where each event generates on average $n_H$ ``child'' events. The process cannot be stable when $n_H > 1$, as events proliferate exponentially with time, and no stationary state can ever be reached. When $n_H < 1$, on the other hand, the average event rate reaches a finite constant that diverges as $(1-n_H)^{-1}$ as $n_H \to 1$.\footnote{The case $n_H=1$ is special and can also reach a stationary state with finite mean intensity when the immigration rate is zero, see \cite{bremaud2001hawkes}, or infinite mean intensity but finite typical (or median) intensity, as shown in \cite{saichev2014neq1}.}
Therefore, for standard Hawkes processes, the stability criterion coincides with the condition that the event rate remains finite.  

As argued by Kanazawa and Sornette in \cite{kanazawa2021exact,kanazawa2021ubiquitous}, non-linear Hawkes processes allow one to combine both excitatory and inhibitory effects, and can describe an even broader range of phenomena. One special class of such non-linear processes, called Quadratic Hawkes Processes (QHP), were introduced and studied in \cite{blanc2017quadratic}. On top of the standard Hawkes feedback, a signed process (price changes in the context of  \cite{blanc2017quadratic}) also contributes to the current activity rate, in a quadratic way (see below for a more precise definition). 
On top of the $n_H$ child events triggered by the Hawkes feedback, the new quadratic feedback contributes to $n_Q$ extra child events. What was shown in \cite{blanc2017quadratic} is that the average event rate now diverges when the {\it total} endogeneity ratio $n_H + n_Q$ reaches unity. From this result, it was concluded that the QHP is only stationary when
$n_H + n_Q < 1$. 

The aim of this note is to show that such a conclusion was too hasty. In fact, we claim that whenever $n_H < 1$ the QHP is always stationary, albeit with a {\it diverging mean intensity} {in the case $n_H + n_Q >1$}. More precisely, the distribution density of the local intensity decays asymptotically as a power-law, with an exponent that becomes smaller than 2 whenever $n_H + n_Q >1$, such that the average intensity diverges. Stationary processes with infinite mean intensity also appeared very recently in the context of non-linear Hawkes processes in Refs. \cite{kanazawa2021exact, kanazawa2021ubiquitous}. Intuitively, the mixed excitatory/inhibitory effects encoded by QHP allows one to avoid the exponential run-away of Hawkes processes when $n_H >1$, while describing a highly intermittent process with divergent mean intensity.

\section{Hawkes Processes: Definition}

\subsection{Linear Hawkes Process}

Hawkes processes were first introduced to model earthquake dynamics \cite{hawkes1971spectra}, in particular the propensity of seismic activity to cluster in time. This same phenomenon is observed in financial markets, where trading activity and volatility tend to cluster in time. 

A Hawkes process $(N_t)_{t\geq0}$ is an inhomogenous Poisson process (meaning that its intensity is time dependent), the intensity of which is defined with the past realisations of the process according to the following equation:\newline
\begin{equation}\label{eq:Hawkes_def}
    \lambda_t= \lambda_\infty + \int_{-\infty}^t\phi(t-u){\rm d}N_u, 
\end{equation}
where $\lambda_t$ is the local intensity, i.e. probability that ${\rm d}N_t$ is equal to 1 is $\lambda_t {\rm d}t$; $\lambda_\infty$ is called the baseline intensity and $\phi(\cdot)$ the influence kernel, from which one obtains the 
endogeneity ratio $n_H$ as the norm of $\phi$:
\begin{equation}
    n_H =||\phi||= \int_\mathbb{R}\phi(u) {\rm d}u <+\infty.
\end{equation}

For an univariate linear Hawkes process to be stable, one needs $n_H<1$. In fact, this is the necessary condition for the mean intensity $\Bar{\lambda}:=\mathbb{E}(\lambda_t)$ to be finite. 

One of the main limitation of Hawkes processes for financial applications is that the stationary distribution of local intensities, $P(\lambda)$, has ``thin tails'' that cannot reproduce the fat tailed distribution of activity/volatility observed in most financial time series. Furthermore, as noted in \cite{blanc2017quadratic}, linear Hawkes processes cannot reproduce the violations of time reversal symmetry observed in financial time series \cite{zumbach2009time}.

\subsection{Quadratic Hawkes}

To overcome the limitations of linear Hawkes processes, Blanc \textit{et al.}~\cite{blanc2017quadratic} introduced a quadratic extension of Hawkes processes, where the intensity is dependent on both past activity (${\rm d}N_t$) and past price returns ${\rm d}P_t := \epsilon_t \psi {\rm d}N_t$, where 
$\epsilon_t = \pm 1$ is an unbiased random sign, independently chosen at each price change, and $\psi$ is the size of elementary price changes. The QHP is now defined as:  
\begin{equation}\label{eq:Lambda_def_QH}
\begin{aligned}
    \lambda_t =& \lambda_{\infty} + \frac{1}{\psi}\int^t_{-\infty}L(t-s){\rm d}P_s\\
    &+\frac{1}{\psi^2}\int^t_{-\infty}\int^t_{-\infty}Q(t-s,t-u){\rm d}P_s{\rm d}P_u,
\end{aligned}
\end{equation}
where $L(\cdot)$ is called the leverage kernel (breaking the ${\rm d}P_t \to - {\rm d}P_t$ symmetry) and $Q(\cdot,\cdot)$ the quadratic kernel. Note that $L$ and $Q$ must be such that the quadratic form in ${\rm d}P_s$ is positive definite, see \cite{blanc2017quadratic}.

Since ${\rm d}P_t^2 = \psi^2 {\rm d}N_t$, a purely diagonal $Q$ (to wit: $Q(t-s,t-u) = \phi(t-s) \delta(s-u)$) recovers the standard Hawkes kernel. New effects arise when considering non-diagonal contributions to $Q$, see below.

Taking the expectation of Eq.~\eqref{eq:Lambda_def_QH} provides an exact equation for the mean intensity $\Bar{\lambda}$, provided it exists. Noting that $\mathbb{E}[\epsilon_s]=0$  and $\mathbb{E}[\epsilon_s \epsilon_u]= \delta(s-u)$, one readily obtains:
\begin{equation}\label{eq:Lambda_QH_mean}
    \Bar{\lambda} = \lambda_{\infty} +  n\Bar{\lambda}; \qquad n:= \int_0^{+\infty} Q(s,s){\rm d}s.
\end{equation}
Hence, $\Bar{\lambda}=\lambda_\infty /(1-n)$ is positive and finite whenever
$n < 1$, but becomes formally negative when $n > 1$ which 
was interpreted in \cite{blanc2017quadratic} as a regime where the QHP becomes non-stationary, in analogy with what happens in the case of standard linear Hawkes processes. As we shall see below, this conclusion is not always warranted. 

\subsection{ZHawkes} 

As an interesting special case that captures the Zumbach effect (i.e. the correlation between future volatility and past trends \cite{zumbach2009time,zumbach2010volatility}), Blanc \textit{et al.}~\cite{blanc2017quadratic} proposed the following ZHawkes specification:
\begin{equation}\label{eq:ZHawkes_K}
    L(s) \equiv 0; \qquad Q(s,u)=\phi(s)\delta(s-u) + z(s)z(u),
\end{equation}
i.e. a quadratic kernel that is diagonal plus a rank-one contribution. In this case, the intensity of the QHP becomes:
\begin{equation}\label{eq:ZHawkes_int}
  \lambda_t=\lambda_\infty+H_t+Z_t^2  
\end{equation}
where
\begin{align*}
    H_t=\int_{-\infty}^t \phi(t-s){\rm d}N_s
\end{align*}
represents the Hawkes component of the intensity whereas 
\begin{align*}
    Z_t=\int_{-\infty}^t z(t-s){\rm d}P_s
\end{align*}represents the trend-induced (Zumbach) component. 
Correspondingly, one can then obtain the endogeneity ratio $n$ (defined in Eq.~\eqref{eq:Lambda_QH_mean}) as the sum of the two terms: the Hawkes endogeneity ratio $n_H:=||\phi||$ and the Zumbach endogeneity ratio $n_Z=||z^2||$. Naively, the stability of the ZHawkes process should read:
\begin{align}
    n = n_H+n_Z<1.
\end{align}
However, our simulations show that one can have a non explosive process when $n\geq1$ provided $n_H<1$. In this case, the QHP is stationary but with an infinite mean intensity $\Bar{\lambda}$. The stability of the QHP for $n_Z+n_H>1$ arises from the fact that the inhibiting realisations of $Z_t$ (corresponding to locally mean-reverting behaviour of the price) are compensating the exciting ones (corresponding to local trends), see the detailed discussion in \cite{kanazawa2021exact}.

In the following, we show the results of our simulations of an univariate ZHawkes process (ZHP) with $n_H=0$ and $n_Z>1$, and an exponential Zumbach kernel $z(\cdot)$. We also simulate the corresponding continuous limit of the ZHawkes process, as worked out in \cite{blanc2017quadratic}, and reinterpret the analytical results of Blanc et al. in the context of the present discussion.

\begin{figure*}[t!]
\centering
\includegraphics[width=\linewidth]{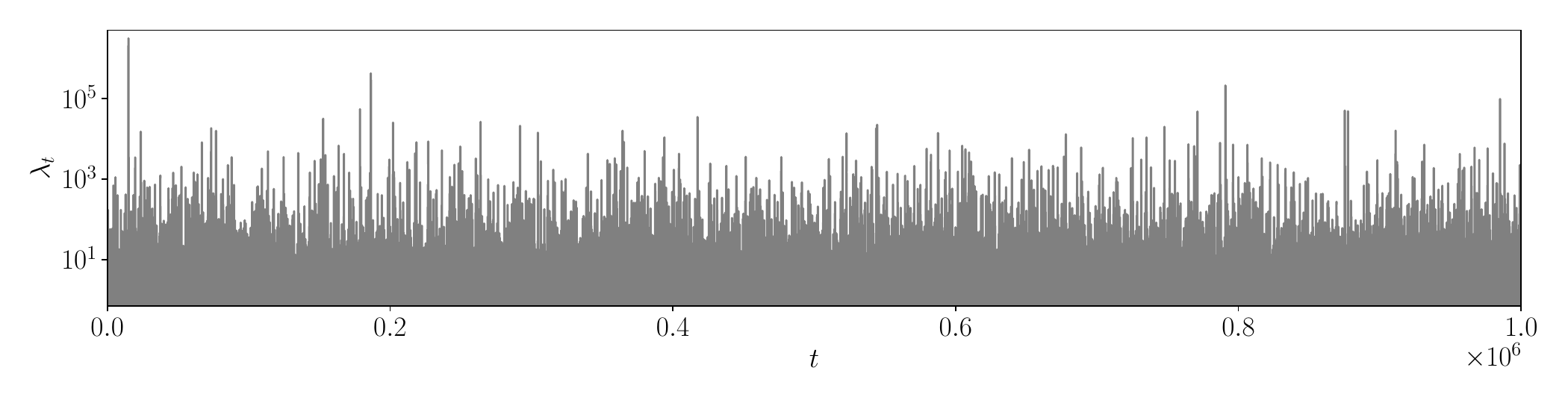}
    \caption{Time Series of the Intensity of a simulated ZHP - Parameters: $n_Z=2$, $T = 10^6$, $\omega=0.03$, $n_H=0$ $\lambda_\infty=0.5$. Note the log-scale on the y-axis.}
    \label{fig:thinningTimeSerie}
\end{figure*}

\section{Univariate 
ZHP: Numerical Results} \label{sec:conclusions}


In this section, we provide evidence that the intensity of a simulated ZHawkes Process with $n_Z>1$ is stationary. In order to simulate a ZHP process, we adapt the thinning algorithm presented by Ogata (1981) in \cite{ogata1981lewis}. The Zumbach kernel $z(\cdot)$ is chosen to be a pure exponential, such that $Z_t$ can be interpreted as an exponential moving average of past returns, i.e. a proxy for the recent trend in prices. More precisely, we set
\[
z(s)=\gamma \, e^{-\omega s},\qquad \gamma:=\sqrt{2n_Z\omega}
\]
with $n_Z=2$, $\omega=0.03$ and a total simulation time of $T=10^6$.  We also set $n_H = 0$, i.e. no Hawkes feedback, and choose the baseline rate to be $\lambda_\infty=0.5$.

\begin{figure}[]
     \centering \includegraphics[width=\linewidth]{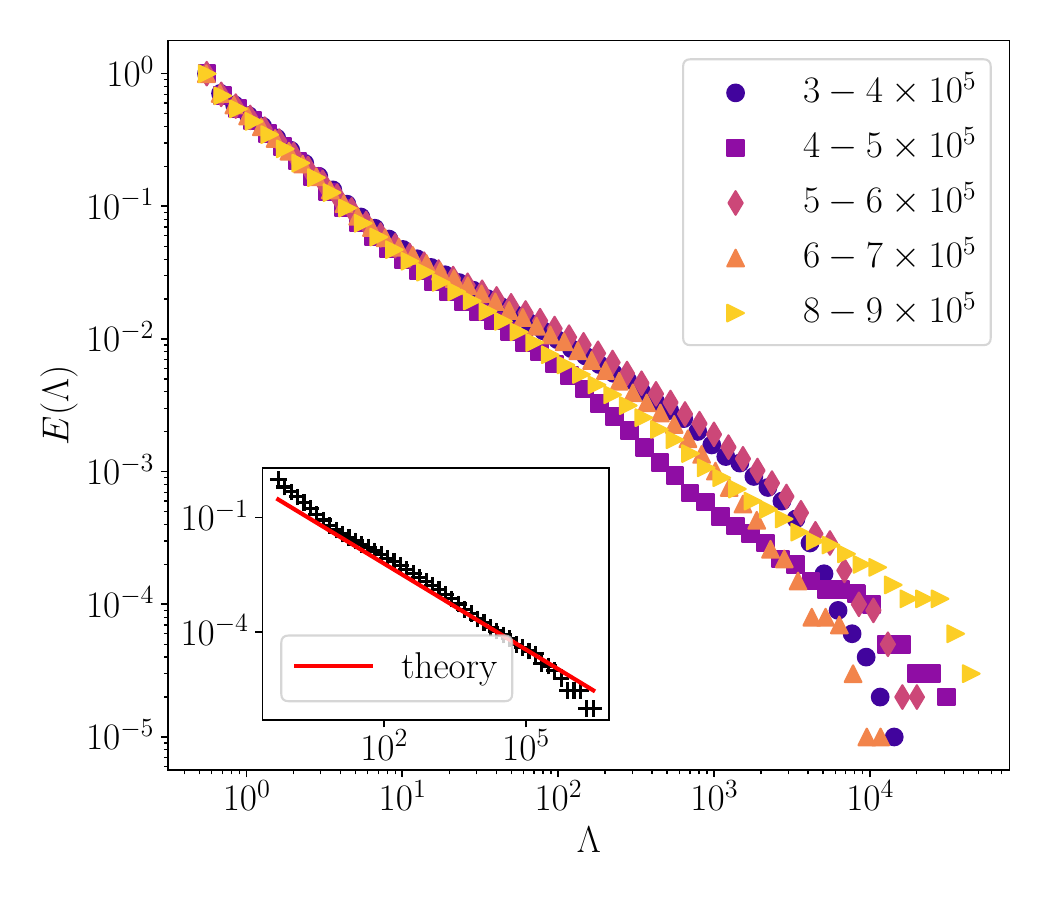}
     \caption{The main figure shows the survival function $E(\Lambda)=\mathbb{P}[\lambda > \Lambda]$ in log-log, for several subperiods $t \in (n \times 10^5, (n+1) \times 10^5)$ of the total simulation, with $n$ indicated in the legend. The results fluctuate somewhat, but there is no systematic trend towards a fatter tail at longer times. The inset shows (in log-log) the survival function for the whole period $10^5 \leq t \leq T= 10^6$, together with the theoretical prediction for the tail of the distribution (red line), as given by Eq. \eqref{eq:power-law}.  Parameters are: $n_Z=2$, $T = 10^6$, $\omega=0.03$, $n_H=0$ and $\lambda_\infty=0.5$.}
     \label{fig:His100000}
\end{figure}

Figure \ref{fig:thinningTimeSerie} represents the whole time series of the simulated intensity, which shows that the process does not explode and looks stationary. More precisely, we find that the survival function of the process $E(\Lambda) := \mathbb{P}[\lambda \geq \Lambda]$ does not significantly evolve with time, see 
Fig. \ref{fig:His100000}. In particular, the distribution does not become significantly ``fatter'' as time increases, as would be expected if the process was on an explosive path. (Note that a formal Kolmogorov-Smirnov test of this statement is not straightforward because the $\lambda_t$'s are correlated in time, see \cite{chicheportiche2011goodness}).
Finally, and most importantly, the empirical distribution function very nicely matches the theoretical prediction of \cite{blanc2017quadratic}, namely:
\begin{equation}\label{eq:power-law}
    E(\Lambda) \underset{\Lambda \gg \lambda_\infty}{\propto} \Lambda^{-\frac{1}{2}(1 +\frac{1}{n_Z})}, \qquad (n_H=0):
\end{equation}
see the inset in Fig.~\ref{fig:His100000}. The expected slope $-3/4$ for $n_Z=2$ is indeed very close to the fitted slope $-0.77$ in the range $\Lambda \in [10^2,10^5]$, beyond which finite size effects become visible. 


\clearpage
\newpage

\onecolumngrid

\begin{figure*}[]
\centering
\includegraphics[width=\linewidth]{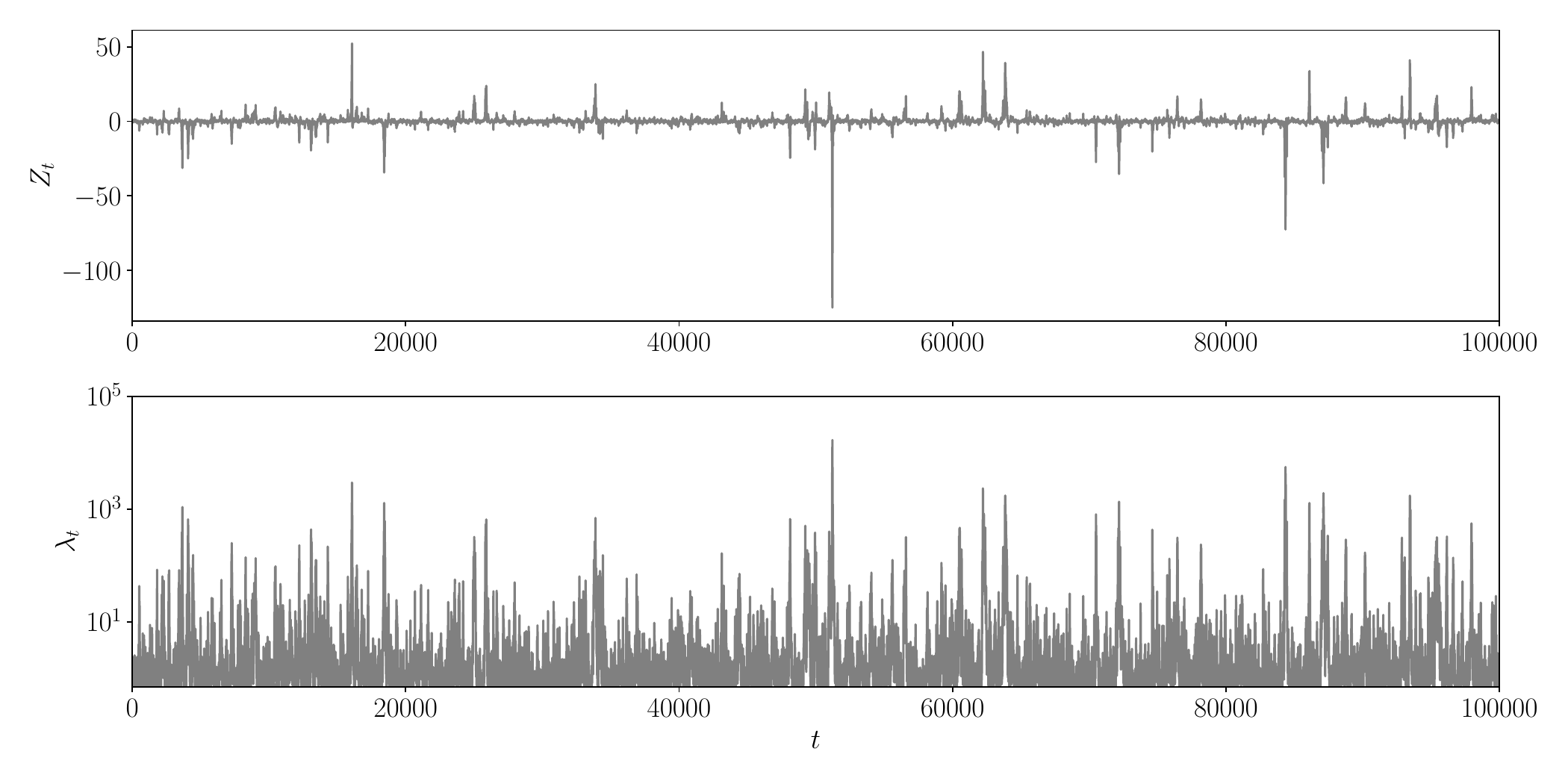}
    \caption{Times Series of the Continuous Time Limit of a ZHawkes process according to Eq. \eqref{eq:continPaths}. Top:  $(Z_t)_t$; Bottom: $\lambda_t=\lambda_\infty+H_t+Z_t^2$ - Parameters: $N=10^5$ $\omega=0.1$ $\beta=1$, $\lambda_\infty=0.5$ $n_H=0.2$ $n_Z=1.5$. Note the log scale on the $y$-axis for $\lambda_t$.}
    \label{fig:timeSeriesContinousSim}
\end{figure*}

\twocolumngrid

\section{Continuous Time Limit of the ZHP}

We now further assume that the Hawkes kernel $\phi(\cdot)$ is also exponential and reads 
\[
\phi(s) = n_H \beta \, e^{-\beta s}.
\]
When the parameters $\omega, \beta$ in the kernels $z(\cdot)$ and $\phi(\cdot)$ are sufficiently small, a continuous time limit of the ZHP was derived in \cite{blanc2017quadratic}. The corresponding two-dimensional SDE reads  
\begin{equation}\label{eq:continPaths}
\begin{cases}
    {\rm d}{H}_t &= \beta \big[-(1-n_H){H}_t+n_H(\lambda_\infty+({Z}_t)^2)\big]{\rm d}t\\
    {\rm d}{Z}_t &= -{\omega}{Z}_t {\rm d}t +{\gamma}\sqrt{\lambda_\infty+{H}_t+({Z}_t)^2}{\rm d}W_t,
\end{cases}
\end{equation}
where ${\rm d}W_t$ is a Wiener noise. Instead of simulating the original ZHP using the thinning method of the previous section, one can simulate the above SDE, with results shown in Fig. \ref{fig:timeSeriesContinousSim}, this time with a non zero Hawkes parameter $n_H=0.2$, and with $\omega=0.1$, $\beta=1$, $\lambda_\infty=0.5$, $n_Z=1.5$; such that $n = 1.7 > 1$ but $n_H < 1$.  The resulting time series of the process $Z_t$ and the intensity $\lambda_t$ are presented in Fig. \ref{fig:timeSeriesContinousSim} and look, again, perfectly stationary, even though the criterion ensuring that $\mathbb{E}[{H}] < + \infty$ and $\mathbb{E}[{Z^2}] < + \infty$, namely $n_H + n_Z < 1$, is violated here.

In fact, the stationary pdf of the two-dimensional process Eq. \eqref{eq:continPaths} was studied in 
\cite{blanc2017quadratic}. It was shown that the tail of the distribution of the intensity $\lambda_t$ is a power-law, given by an extension of Eq. \eqref{eq:power-law}:
\begin{equation}\label{eq:power-law2}
    E(\Lambda) \underset{\Lambda \gg \lambda_\infty}{\propto} \Lambda^{-\frac{1}{2}(1 +\frac{1}{n_Z(1+a)})}, 
\end{equation}
where $a$ can only be computed in some limits:
\begin{equation} 
a \approx \frac{n_H}{1 - n_H}\left[1 - \chi \frac{1 - n_H - n_Z}{(1-n_H)^2}\right] \quad (\chi = \frac{2 \omega}{\beta} \to 0),
\end{equation}
and
\begin{equation} 
a \approx \frac{n_H}{\chi(1 - n_Z)} \quad (\chi = \frac{2 \omega}{\beta} \to \infty).
\end{equation}
Note that the latter expression is in fact valid for arbitrary $\chi$ when $n_H \to 0$, provided $n_H \ll \chi(1-n_Z)$. In particular, $a=0 $ when $n_H=0$, recovering Eq. \eqref{eq:power-law}.

The conclusion is that the ZHP, at least in the continuum limit, always reaches a stationary distribution when $n_H < 1$, albeit the tail of the distribution of $\lambda_t$ is a power-law, with an exponent that becomes smaller than unity when $n_H + n_Z > 1$, i.e. leads in that case to a divergent mean intensity.

\section{Conclusion} \label{sec:conclusion}

In this work, we have revisited the properties of Quadratic Hawkes processes in the strong feedback regime. Based on numerical simulations and analytical results, we have argued that a new regime exists, where the process reaches a stationary state with an infinite mean intensity. Such a regime is absent for standard (linear) Hawkes processes: the stability of the process requires the mean intensity to be finite, except in the critical case \cite{saichev2014neq1}. As argued by Kanazawa \& Sornette \cite{kanazawa2021exact, kanazawa2021ubiquitous}, non-linear Hawkes processes allow for a rich phenomenology, with inhibitory and excitatory effects that can balance each other in a subtle way, resulting in a highly fluctuating, but non-explosive process, for which they provide several other examples. 

QHP naturally lead to a power-law tail distribution for the local intensity, with an exponent that can become less than unity, in which case the mean intensity diverges. The resulting price process then converges to a L\'evy stable process with an infinite variance, in spite of the fact that elementary price changes are strictly bounded (and, in our example, equal to $\pm \psi$). As such, this regime is not directly relevant to financial markets, since price returns, although fat-tailed, have a finite variance. Notwithstanding, we believe  that the possibility of creating L\'evy stable random walks based on a self-exciting mechanism is interesting in itself, and may have applications in other fields. 

Finally, our analysis is far from mathematically rigorous. Although the results of \cite{blanc2017quadratic} for the continuous time version of the Z-Hawkes process are suggestive (see section IV), a more formal proof of the stationarity of Quadratic Hawkes processes with general kernels and in the infinite mean intensity regime would be welcome. An extension of the present discussion to multivariate Quadratic Hawkes processes is under way.

\section*{Acknowledgements} \label{sec:acknowledgements}

We want to thank Jérôme Garnier-Brun, Samy Lakhal and Rudy Morel for very fruitful discussions. \newline

This research was conducted within the
Econophysics \& Complex Systems Research Chair, under the aegis of the Fondation du Risque, the Fondation de l’Ecole polytechnique, the Ecole polytechnique and Capital Fund Management.

\bibliographystyle{apsrev4-1}
\bibliography{biblio}

\onecolumngrid

\end{document}